\documentclass[aps,pra,floatfix,preprint]{revtex4-1}
\usepackage{amsmath,amssymb,amsfonts,bbm,graphicx,hyperref}
\usepackage[utf8]{inputenc}
\usepackage[english]{babel}
\usepackage{graphicx}
\usepackage{amsmath,times,amssymb}
\usepackage{amsfonts}
\usepackage{mathrsfs}
\usepackage{physics}
\usepackage{changes}
\def\Tr{\hbox{Tr}}
\begin{document}
\title{Isospectral oscillators as a resource for quantum information processing}
\markboth{A. Chabane et al.}{Isospectral oscillators as a resource for quantum information processing}
\author{Abdelatif Chabane$^1$, Sidali Mohammdi$^1$, Abdelhakim Gharbi$^1$, Matteo G. A. Paris$^2$}
\affiliation{${}^1$Universit\'e de Bejaia, Facult\'e des Sciences Exactes, Laboratoire de Physique Th\'eorique, 06000 Bejaia, Alg\'erie \\ 
${}^2$Dipartimento di Fisica {\em Aldo Pontremoli}, Universit\`a degli Studi di Milano, I-20133 
	Milano, Italy}
	\date{\today}
%%%%%%%
\begin{abstract}
We address quantum systems isospectral to the harmonic oscillator, as those found within the framework of supersymmetric quantum mechanics, as potential resources for continuous variable  quantum information. These deformed oscillator potentials share the equally spaced energy levels of the shifted harmonic oscillator but differ significantly in that they are non-harmonic. Consequently, their ground states and thermal equilibrium states are no longer Gaussian and exhibit non-classical properties. We quantify their non-Gaussianity and evaluate their non-classicality using various measures, including quadrature squeezing, photon number squeezing, Wigner function negativity, and quadrature coherence scale. Additionally, we employ quantum estimation theory to identify optimal measurement strategies and establish ultimate precision bounds for inferring the deformation parameter. Our findings prove that quantum systems isospectral to the harmonic oscillator may represent promising platforms for quantum information with continuous variables. In turn, non-Gaussian and non-classical stationary states may be obtained and 
these features persist at non-zero temperature.
\end{abstract}
\maketitle
%%%%
\section{Introduction}
In quantum information processing, the unique features of continuous variable (CV) quantum systems make them valuable for certain tasks\cite{CVQI,andersen2010continuous,cerf2007quantum,ferraro2005gaussian,giedke2002characterization,navarrete2015introduction,killoran2019continuous,lloyd1999quantum,van2002quantum,menicucci2006universal,ukai2010universal,flammia2009optical,serafini2023quantum}. In particular, CV systems may be more robust against certain types of noise and de-coherence compared to discrete variable (DV) systems \cite{cvn1,cvn2}. Additionally, CV systems, especially those realized on optical platforms, are experimentally accessible and can be implemented with existing technology. Finally, CV quantum computing involves fault tolerance \cite{menicucci2014fault} and error correction strategies \cite{braunstein1998error} that differ from those used in DV systems and may be less expensive in term of resources.

In CV quantum information, Gaussian states play an important role for their practical advantages, including simplicity in generation, detection, and manipulation. However, Gaussian states have a positive Wigner function\cite{hudson1974wigner,soto1983wigner}, making them of little use in several tasks including quantum simulations\cite{bartlett2002efficient} and computation \cite{ohliger2010limitations}. To overcome these limitations, non-Gaussian states have been widely  investigated\cite{walschaers2021non,kitagawa1987non,molmer2006non,olivares2003teleportation,adnane2019quantum,adnane2019quantum2,genoni2010quantifying} and it has been shown that they can offer advantages and open up new possibilities in several tasks in quantum information science. In turn, non-Gaussian states often exhibit more pronounced non-classical features compared to Gaussian states, thus providing more effective quantum resource for quantum teleportation \cite{olivares2003teleportation,takeda2012quantum}, quantum key distribution \cite{singh2021non,hu2020continuous,leverrier2011continuous,guo2019continuous,huang2013performance}, and quantum computation \cite{niset2009no,knill2001scheme,lloyd1999quantum,adesso2009optimal,sabapathy2017non,cerf2005non},  
being also relevant for the detection of gravitational waves \cite{ligo2011gravitational,aasi2013enhanced,acernese2019increasing,tse2019quantum}.
Those quantum non-Gaussian states are more challenging to generate, manipulate, and detect experimentally compared to Gaussian states, as their generation 
generally involves nolinearity, conditional measurements, or quantum state engineering operations \cite{lloyd1999quantum,walschaers2020practical,lvovsky2020production}, as for example, adding/subtracting photons\cite{ourjoumtsev2006generating,molmer2006non,wenger2004non,zavatta2008subtracting,averchenko2016multimode,parigi2007probing}, which can induce and enhance the amount of non-classicality in an arbitrary quantum state\cite{chuong2023enhancement,malpani2020impact}.

In this scenario, it would be extremely convenient to have a cheaper source of simultaneously non-Gaussian/non-classical states to exploit their properties for CV quantum information processing without dealing with non-linear processes, ancillary systems, and specialized detection methods. 
This is precisely the scope of this paper. 

Using tools from super-symmetric quantum mechanics (SUSYQM), we consider systems described by deformed oscillator potentials, however strictly isospectral to the shifted harmonic oscillator (SHO) \cite{ssqm1,iso0,iso1,iso2,iso3,iso4}. 
This fact makes them suitable candidates for CV quantum information, since 
the potentials are not harmonic, and the ground states, as well as states in thermal equilibrium, are no longer Gaussian and exhibit non-classical properties.
Motivated by these facts, we set ourselves the task to investigate these states by quantifying their non-Gaussianity (nonG) through quantum relative entropy, and their non-classicality (nonC) using quadrature squeezing, photon number squeezing, quadrature coherence scale and Wigner function negativity. 
Finally, we address the characterization of isospectral SHO potentials, and 
seek for the optimal strategy to infer the deformation parameter, using
tools from local quantum estimation theory \cite{holevo2011probabilistic,helstrom1969quantum,paris2009quantum,albarelli2020perspective}. In this procedure, the figure of merit is the so-called quantum Fisher information, which provides a quantitative measure of the information about a parameter, which may be extracted by measurements performed on a family of quantum states. 

The paper is structured as follows. In Section \ref{s:pre}, we briefly provide an introduction to isospectral potentials in one dimensional SUSYQM and report the properties of non-Gaussianity and non-classicality as well as the possible measures to quantify these quantum resources, also review the ideas and the methods of local quantum estimation theory. In Section \ref{s:isoho}, we give the family of isospectral potentials for the shifted harmonic oscillator (SHO) with its eigenfunctions spectrum. In Section \ref{s:nongnonc}, we address the use of non-Gaussianity and non-classicality measures by taking into account two different families of stationary states, ground and thermal states, relevant to SHO isospectral potentials. In Section \ref{s:qet}, we evaluate the quantum Fisher information for the chosen states in order to infer the deformation parameter characterizing the 
SHO isospectral potentials, then we discuss results. Section \ref{s:out} closes the 
paper by some concluding remarks.

\section{Preliminaries}\label{s:pre}
In this Section, we review the main tools used in this paper: the 
notion of isospectral potentials in supersymmetric quantum mechanics, the
quantification of  non-Gaussianity and nonclassicality, and finally the 
formalism of quantum estimation.
\subsection{Isospectral potentials in supersymmetric quantum mechanics}
\label{iso section}
{The formalism of supersymmetry allows for the construction, given some potential $V(x)$, of a family of potentials that have the same spectrum, and the same reflection and transition coefficients as the initial potential. In what follows we briefly review the method to generate a one-parameter isospectral family for a given potential. The ground state energy of the Hamiltonian need to be zero in order for the SUSYQM formalisme to hold, this can be achieved without affecting the physics of the system by shifting the potential of the Hamiltonian.}
{Let us first introduce the partner potentials $V^{(1)}(x)$ and $V^{(2)}(x)$ in terms of the superpotential $W(x)$}
\begin{equation}
{V^{(1)}(x)}={W^{2}}(x)-\frac{\hbar}{\sqrt{2m}}W^{\prime}(x)
\end{equation}%
\begin{equation}
{V^{(2)}(x)}={W^{2}}(x)+\frac{h}{\sqrt{2m}}W^{\prime}(x).
\end{equation}
One can factorize the Hamiltonien in two different ways by introducing the operators $A$ and ${A^{\dagger}}$
\begin{align}
A&=\frac{\hbar}{\sqrt{2m}}\frac{d}{dx}+W(x)\\
A^{\dagger}&=-\frac{\hbar}{\sqrt{2m}}\frac{d}{dx}+W(x)
\end{align}
We then write the corresponding partner Hamiltonians
\begin{align*}
H^{(1)}  &  ={A^{\dagger}A}=-\frac{{{\hbar^{2}}}}{{2m}}\frac{{{d^{2}}}}%
{{d{x^{2}}}}+{V^{(1)}}(x)\\
{H^{(2)}}  &  ={A}{A^{\dagger}}=-\frac{{{h^{2}}}}{{2m}}\frac{{{d^{2}}}}%
{{d{x^{2}}}}+{V^{(2)}}(x).
\end{align*}
It is well established, when supersymmetry is unbroken, that the Hamiltonian
$H^{(2)}$ has the same spectrum of eigenvalues as $H^{(1)}$, except for
missing the ground-state energy of $H^{(1)}$:
\begin{equation}
E_{n}^{(2)}=E_{n+1}^{(1)}.
\end{equation}
Also, the eigenfunctions of the Hamiltonian ${H^{(2)}}$ are related to those
of ${H^{(1)}}$ by:
\begin{equation}
\psi_{n}^{(2)}(x)=\frac{A}{\sqrt{E_{n+1}^{(1)}}}\psi_{n+1}^{(1)}(x)
\end{equation}
Consequently, we can reconstruct all the eigenfunctions of ${H^{(1)}}$ from
those of ${H^{(2)}}$ except for the ground state
\begin{equation}
\psi_{n+1}^{(1)}(x)=\frac{{{A^{\dagger}}}}{\sqrt{E_{n}^{(2)}}}\psi_{n}^{(2)}(x).
\end{equation}
It is worth noting that the superpotential $W(x)$ is related to the ground
state wave function ${\psi_{0}^{(1)}(x)}$ by
\begin{equation}
W(x)=-\frac{\hbar}{\sqrt{2m}}\frac{d}{{dx}}\ln\left(  {\psi_{0}^{(1)}%
(x)}\right)  .
\end{equation}
Using Darboux procedure, { one can} construct a one-parameter family of strictly isospectral potentials ${{\widehat{V}}^{(1)}(\lambda\,;x)}$, all having the same eigenvalues, to a given initial potential ${V^{(1)}(x)}$ having at least one bound state.  By introducing the function%
\begin{equation} \label{a1}
I(x)=\int_{{-\infty}}^{{x}}{{\left[ {\psi_{0}^{(1)}(x{'})}\right]  }^{2}dx{'},}%
\end{equation} 
the isospectral potentials ${{\widehat{V}}^{(1)}(\lambda\,;x)}$ are obtained 
by the formula%
\begin{equation} \label{a2}
{{\widehat{V}}^{(1)}(\lambda\,;x)}=
{V^{(1)}(x)}%
- \frac{\hbar^2}{m} \frac{{{d^{2}}}}{{d{x^{2}%
}}}\ln\left(  {\frac{1}{\lambda}+\frac{\sqrt{2m}}{\hbar}I(x)}\right)  
\end{equation}
where the second term is an anharmonic term that encodes the nonlinear features of these oscillators. Notice that we denote the parameter of the isospectral family by  $1/\lambda$ instead of $\lambda$, as used in literature, in order to better track graphically the asymptotic behavior of our potentials \cite{isocs}.

These potentials have been used to construct explicit solutions of  non-linear evolution equations.\cite{iso2} and have been used to calculate resonances 
in a three-dimensional finite square well \cite{isoex}. The eigenvalues of  ${{\widehat H}^{(1)}(\lambda\,;x)}$ and $H^{(1)}$ are the same.
The eigenstates of ${{\widehat H}^{(1)}(\lambda\,;x)}$ are unitary transforms of those of $H^{(1)}$. The ground state wave functions $\phi_0(x;\lambda)$ may be written in term of ${\psi_{0}^{(1)}(x)}$, the ground state eigenstate of the original Hamiltonian ${H^{(1)}(x)}$, as follows
\begin{align} \label{a3}
\phi_0(x;\lambda)& =\frac{\sqrt{1 + \lambda \frac{\sqrt{2m}}{\hbar}}}{1 + \lambda I(x) \frac{\sqrt{2m}}{\hbar}}\,
\psi_{0}^{(1)}(x),
\end{align}
whereas the excited eigenstates of the isospectral Hamiltonians ${{\widehat{H}}^{(1)}(\lambda\,;x)}$ are given in terms of both, the excited eigenstates of the  states original Hamiltonian ${H^{(1)}(x)}$ and their corresponding eigenvalues, by the expression
\begin{align*} \label{a4}
\phi_{1+n}(x;\lambda)&=  \left[ 1+\frac{1}{{E_{n+1}^{(1)}}%
}\left(  {\frac{{I^{\prime}(x)}}{{\frac{1}{\lambda}+\frac{\sqrt{2m}}{\hbar
}I(x)}}}\right) \left(  {\frac{\hbar}{\sqrt{2m}}\frac{d}{dx}+W(x)}\right)\right]
  \psi_{n+1}^{(1)}(x)
\end{align*}
It should be noted that the eigenfunctions $\phi_n(x;\lambda)$ are square-integrable if and only if $\lambda>-\frac{\hbar}{\sqrt{2m}}$. This is the case we are going to consider in this work.
\subsection{Quantification of non-Gaussianity}
The non-Gaussianity (nonG) of a CV quantum state $\rho$ may be quantified 
in term of the statistical distinguishability of the state from a reference 
Gaussian state $\rho_{G}$, chosen to have the same first-moment vector $E\left[\rho_{G}\right]$ and the same covariance matrix $\sigma\left[\rho_{G}\right]$ of $\rho$ (see Appendix \ref{Apx-A}). In particular, the distinguishability is quantified 
using the quantum relative entropy 
\begin{equation}
\delta\lbrack\rho]=S\left(  {\rho\parallel{\rho_{G}}}\right)  =S\left(
{{\rho_{G}}}\right)  -S\left(  \rho\right)\,,
\end{equation}
where $S\left(  \rho\right)  =-\hbox{Tr}\left[  {\rho\log\,\rho\,}\right]$ 
is the Von-Neumann entropy of $\rho$ and for a single-mode Gaussian states 
\begin{equation}
S\left(  {{\rho_{G}}}\right)  =h\left(  \sqrt{\det\,\sigma\left[  \rho\right] }\right) ,
\end{equation}
where $h(x)$ is a function given by:
\begin{equation}
h(t)=(t+\frac{1}{2})\ln(t+\frac{1}{2})+(t-\frac{1}{2})\ln(t-\frac{1}{2})
\end{equation}
Overall, we have the the nonG $\delta[\rho]$ of a CV quantum state $\rho$ may be expressed as 
\begin{equation}
\delta\left[  \rho\right]  =h\left(  \sqrt{\det\,\sigma\left[  \rho\right] }\right)  -S(\rho)
\label{nong}
\end{equation}
%%%
\subsection{Non-classicality measures}
\label{noncapproch}
Non-classical states are conventionally defined as those that cannot be expressed as a statistical mixture of coherent states. This definition is grounded in physical principles and encompasses phenomena that may or may not be observable in experiments. However, in quantum information processing, other characteristics beyond non-classicality can be crucial for achieving quantum advantage. This has led to the development of more restrictive notions of non-classicality, tailored to specific applications \cite{lee1991measure,lee1992moments,paris2001robust,dodonov2002nonclassical,marian2002quantifying,dodonov2003classicality,shchukin2005nonclassical,ferraro2012nonclassicality,glauber1963coherent,kenfack2004negativity,mari2011directly,de2019measuring,hertz2020quadrature,hertz2022decoherence}. Evaluating quantum states across these various criteria is essential to assess their potential use in quantum technology. Although it is regrettable that all these measures are referred to as non-classicality measures, this reflects the current state of the literature, and we will not introduce new terminology here.

In this work, we aim to provide a realistic qualitative and quantitative characterization of both the ground states and the equilibrium states of SHO isospectral potentials. To achieve this, we employ various measures and criteria for non-classicality, including those based on quadrature squeezing and photon number squeezing, which serve as necessary but not sufficient conditions for fully establishing the non-classical nature of a quantum state. Additionally, we utilize quantitative measures such as the negativity of the Wigner function and the quadrature coherence scale (QCS) to further refine our analysis.

\subsubsection{Quadrature and photon number squeezing}
Coherent states are minimum uncertainty states, i.e. with uncertainties that are equal for position $X$ and momentum $P$ operators. Similarly squeezed states are minimum uncertainty states but they exhibit less uncertainty in one quadrature and increased fluctuations in the other one. They are non-classical states since this is not compatible with being a mixture of coherent states.
For any quantum state, the pair of canonical operators $\left( X,P\right) 
$ obey the Heisenberg uncertainty
relation:
\begin{equation}
\Delta X^{2}\Delta P^{2}\geqslant \frac{{{\hbar^{2}}}}{4}%
\end{equation}
where ${\Delta X}^{2}=\left\langle
X^{2}\right\rangle -\left\langle X\right\rangle ^{2}$ and $
{\Delta P}^{2}=\left\langle
{P^{2}}\right\rangle -\left\langle {P}\right\rangle ^{2}$
are the variances of the position $\widehat{{X}}$ and momentum $\widehat{{P}}$ operators respectively. also, the shot noise variance is expressed by the quantity $\dfrac{\hbar}{2}$.
The state is said to be squeezed if the variance according to a quadrature is compressed (less than $\hbar/2$). Once we get this squeezing signature, we can say that our state has non-classical features since squeezing implies the singularity of the P-function. Squeezing is thus a sufficient condition for non-classicality (but not necessary)\cite{agarwal2012quantum}.

Similarly, one speaks of photon number squeezing for those states which satisfy the relation ${\left( {\Delta {\widehat n}}\right)^{2} < {{\left\langle {\widehat n} \right\rangle }}}$ where $\widehat n = n \left| n \right\rangle \left\langle n \right|$ is the photon number operator. This is usually quantified by means of lthe Fano factor
\begin{equation}
{\cal F}[\rho] = \frac{{{{\left( {\Delta \hat n} \right)}^2}}}{{\left\langle {\hat n} \right\rangle }} = \frac{{{{\left\langle {{{\hat n}^2}} \right\rangle } } - {{\left\langle {\hat n} \right\rangle }}^2}}{{{{\left\langle {\hat n} \right\rangle }}}}\,.
    \label{Ff}
\end{equation}
If ${\cal F}[\rho]<1$, the photon number distribution is sub-Poissonian and the state is non-classical since this condition is incompatible with having a positive Glauber $P$-function. Photon number squeezing is a sufficient condition for non-classicality but it is not a necessary one. 
\subsubsection{Wigner negativity}
A celebrated phase-space description of non-classicality in single mode quantum oscillators is the volume occupied by the negative part of the Wigner function. 
Since the Wigner function $W(x,p)$ of a quantum state $\rho$
\begin{equation}
W(x,p) = \frac{1}{{\pi \hbar }}\int\limits_{ - \infty }^{ + \infty } {\left\langle {x + y} \right|\rho \left| {x - y} \right\rangle {e^{ - \frac{{2ipy}}{\hbar }}}} dy
\label{WF}
\end{equation}
is a normalized (not positive-definite )quasi-probability distribution in the phase space, the \textit{W-nonclassicality} may be written as 
\begin{equation}
\nu[\rho] = \left( {\iint {dx\,dp\,\left| {W(x,p)} \right|}} \right) - 1
\label{NC}\,.
\end{equation}
%%%
\subsubsection{Quadrature coherence scale (QCS)}
The term \textit{coherences} usually denote the off-diagonal elements 
$\rho_{aa'} = \langle a |\rho| a' \rangle$ of the density matrix in the eigenbasis of a given observable $A$. To assess the overall coherence of the state one may introduce the $A$-coherence scale (squared) of $\rho$ via
\begin{equation}
{\cal C}_A^2\left( \rho  \right) = \frac{1}{{\cal P}}\Tr\left[ {\rho ,A} \right]\left[ {A,\rho } \right] = \frac{1}{{\cal P}}{\int (a - {a'})^2|\rho ( a,{a'})| ^2 da d{a^{'}}}
\label{QCSA}
\end{equation}
where ${\cal P}\left[ \rho  \right] = \Tr\left[ {{\rho ^2}} \right]$ is the purity of the state $\rho$. For a CV system, characterized by the pair of conjugate quadratures $(X,P)$, the QCS is given by 
\begin{equation}
{{\cal C}^2}\left( \rho  \right) = \frac{1}{2}\left[ {{\cal C}_X^2\left( \rho  \right) + {\cal C}_P^2\left( \rho  \right)} \right]
\label{QCS} 
\end{equation}
and quantifies how far from the diagonal the off-diagonal coherences of all quadratures $(X,P)$ lie. A large ${\cal C}(\rho)$ implies that for the pair $(X,P)$, at least one has a large coherence scale.  When the QCS is small, i.e. ${\cal C}(\rho)\leq 1$, the states are close to classical states, whereas for a large QCS, i.e., ${\cal C}(\rho)\gg1$, the states are strongly non-classical, and very sensitive to environmental decoherence.
\subsection{Local quantum parameter estimation }
Many quantities of interest in quantum physics are not accessible by direct measurements because they do not correspond to quantum observables. One can cite temperature, optical phase, and the parameter $\lambda$ characterizing a family of isospectral potentials. To resolve this issue and assign these quantities a value, we have recourse to indirect measurements within the quantum estimation theory (QET) which provides tools to infer the value of an unknown parameter with the optimal possible precision in the quantum limit by finding the best measurements and the best states to achieve this task.
In a classical estimation problem, we look for an estimator defined as a function of a finite set of empirical data $\{x_{1},x_{2},...,x_{m}\}$ to the set of possible values of the parameter
$
\hat{\lambda}=\hat{\lambda}\left(  x_{1},x_{2},...,x_{m}\right)$.
An unbiased estimator is said to be optimal if it saturates the so-called Cramèr-Rao bound (CRB)
\begin{equation}
\hbox{Var}(\hat{\lambda})\geqslant\frac{1}{{MF(\lambda)}},
\end{equation}
which establishes a lower bound on the variance $\hbox{Var}(\hat{\lambda})=E_{\lambda
}\left[  \left(  \hat{\lambda}\left(  x\right)  -\lambda\right)  ^{2}\right]
${ of any estimator}, $M$ is the number of times
the measurement is repeated, and $F(\lambda)$ is the Fisher Information (FI):
\begin{equation}
F(\lambda)=\int{dx\frac{1}{{p(x|\lambda)}}}{\left(  {\frac{{\partial
p(x|\lambda)}}{{\partial\lambda}}}\right)  ^{2},}%
\end{equation}
here $p(x|\lambda)$ denotes the conditional probability of obtaining the
measurement outcome value $x$ when the parameter has the value $\lambda.$ the
Fisher Information $F(\lambda)$ can be viewed as a measure of the amount of
information carried $p(x|\lambda)$ with respect to $\lambda.$
In quantum mechanics, the conditional probability $p(x|\lambda)$ is given by
the Born rule $p(x|\lambda)=\Tr\left[  \rho_{\lambda}\hat{\Pi}_{x}\right]  $,
where $\rho_{\lambda}$ belongs to a family of quantum states $\left\{
\rho_{\lambda}\right\}  $ depending on the parameter $\lambda$, and forming the
quantum statistical model under investigation. $\hat{\Pi}_{x}$ is an element
of the positive operator-valued measure (POVM) representing the measurement
with the requirement $\sum\limits_{x}\hat{\Pi}_{x}={\normalsize 1}$. Then, FI
takes the form
\begin{equation}
F(\lambda)=\int{dx\frac{{\left(  {\partial}_{\lambda}\Tr\left[  \rho_{\lambda
}\hat{\Pi}_{x}\right]  \right)  ^{2}}}{\Tr\left[  \rho_{\lambda}\hat{\Pi}%
_{x}\right]}}.
\label{FI}
\end{equation}
By introducing the Hermitian operator $L_{\lambda}$, known as the symmetric logarithmic derivative (SLD), which is the solution of the equation
$\partial_\lambda\rho_\lambda =\frac12 (L_\lambda \rho_\lambda +\rho_\lambda L_\lambda)$
one shows that FI can be maximized over all possible generalized measures POVMs by the so-called quantum Fisher information (QFI) $H\left(\lambda\right)$:
\begin{equation}
H\left(  \lambda\right)  =\Tr\left[  \rho_{\lambda}L_{\lambda}^{2}\right]  \geq
F(\lambda),
\end{equation}
this leads straightforwardly to the definition of the quantum Cram\'{e}r-Rao
bound (QCRB)%
\begin{equation}
\hbox{Var}(\widehat{\lambda})\geq\frac{1}{M{\,H(\lambda)}}%
\label{QCRB}
\end{equation}
we notice that the QFI depends only on the quantum state $\rho_{\lambda}$, and
determines the ultimate bound to precision of any estimation strategy for
${\lambda.}$

By writing the quantum state $\rho_{\lambda}$ in its eigenbasis%
$
{{\rho_{\lambda}}=\sum_{n}{{\rho_{n}}(\lambda)\left\vert {{\psi_{n}}(\lambda
)}\right\rangle \left\langle {{\psi_{n}}(\lambda)}\right\vert ,}}%
$
where ${{\left\vert {{\psi_{n}}(\lambda)}\right\rangle }}$ is an eigenvectors
of ${{\rho_{\lambda}}}$ and ${{{\rho_{n}}(\lambda)}}$ the corresponding
eigenvalue, we then obtain an explicit expression of the QFI given by%
\begin{align}
H(\lambda)&=\sum_{n}{\frac{{{{\left(  {{\partial_{\lambda}}{\rho_{n}}(\lambda
)}\right)  }^{2}}}}{{{\rho_{n}}(\lambda)}}}\label{QFIM}\\
&+2\sum_{n\neq m}{{\frac{{{{\left(
{{\rho_{n}}(\lambda)-{\rho_{m}}(\lambda)}\right)  }^{2}}}}{{{\rho_{n}}%
(\lambda)+{\rho_{m}}(\lambda)}}{{\left\vert \left\langle {{{\psi_{m}}%
(\lambda)}}\right.  \left\vert {{\partial_{\lambda}{\psi_{n}}(\lambda)}%
}\right\rangle \right\vert }^{2}.}}}\notag
\end{align}
The first term represents the classical Fisher information of the probability
distribution $\{p_{k}\}$, whereas the second term contains the truly quantum
contribution to the information accessible to a quantum system.
For a generic family of pure states ${{\rho_{\lambda}}=\left\vert {{\psi
}(\lambda)}\right\rangle \left\langle {{\psi}(\lambda)}\right\vert }${,} the
QFI reduces to the following form:
\begin{equation}
H(\lambda)=4\left[  \left\langle {{\partial_{\lambda}{\psi}(\lambda)}}\right.
\left\vert {{\partial_{\lambda}{\psi}(\lambda)}}\right\rangle {+{{\left(
\left\langle {{\partial_{\lambda}{\psi}(\lambda)}}\right.  \left\vert {{{\psi
}(\lambda)}}\right\rangle \right)  }^{2}}}\right].
\label{QFIP}
\end{equation}
Note that in a given quantum estimation problem, a quantum measurement is optimal if the condition ${F(\lambda)=}H(\lambda)$ is met. An estimator is said to be efficient if the corresponding variance saturates the quantum Cram\'{e}r-Rao inequality. For several relevant quantity in CV quantum information optimal estimation may be indeed achieved, including entanglement
\cite{genoni2008optimal,brida2010experimental}, discord 
\cite{PhysRevA.85.064104,virzi2019optimal}, loss/dissipation \cite{monras2007optimal,invernizzi2011optimal}, and temperature itself
\cite{paris2015achieving}.
\section{Isospectral potentials of the shifted harmonic oscillator}\label{s:isoho}
\begin{figure}[h!]
\centering
\includegraphics[width=0.8\columnwidth]{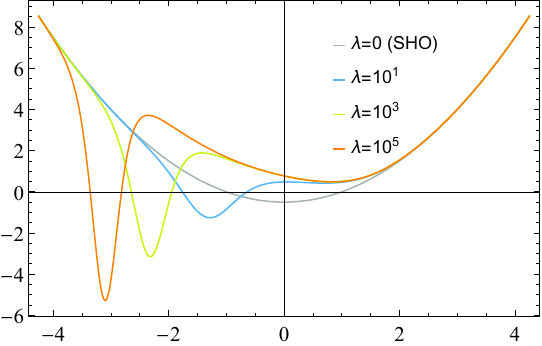} 
\includegraphics[width=0.4\columnwidth]{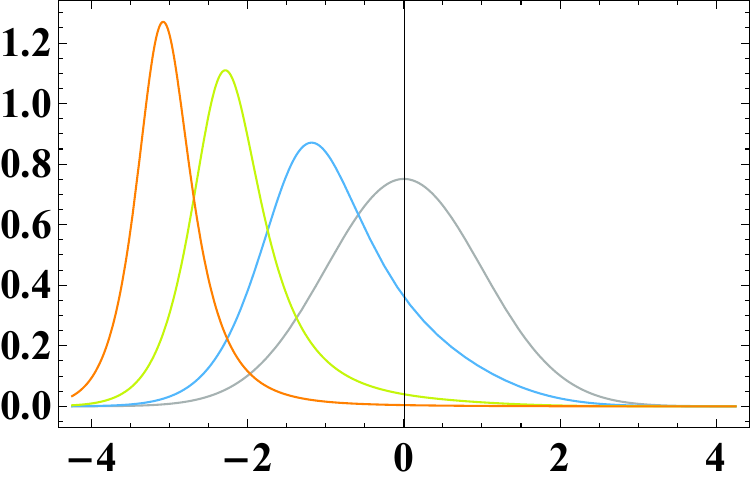}
\includegraphics[width=0.4\columnwidth]{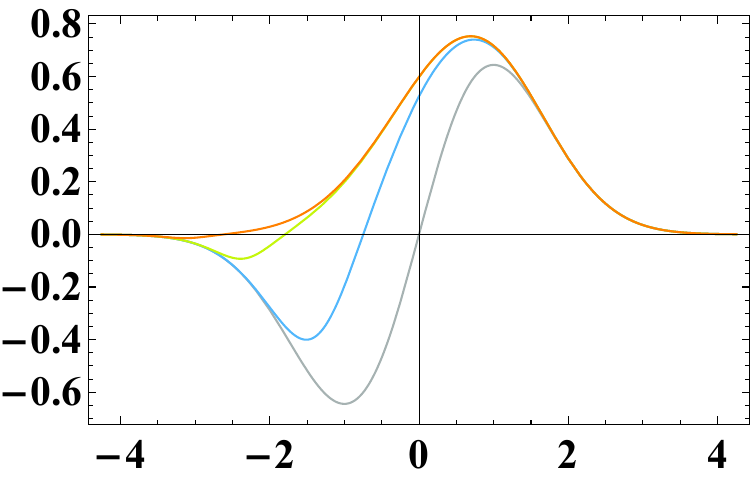}
\caption{In the top panel, we show the isospectral SHO potentials ${{\widehat V}^{(1)}(\lambda\,;x)}$ for $\lambda=0$ (Gray), $\lambda=10$ (blue), $\lambda=10^{3}$ (green), $\lambda=10^{5}$ (orange). In the lower left panel, we show the wave functions of the ground state $\phi_0(x;\lambda)$ and, in the lower right panel, those of the first excited state $\phi_1(x;\lambda)$. The color code is the same of the above panel.}
\label{isoSHO}
\end{figure}

In this Section, we discuss the properties of the supersymmetric isospectral potentials to 
the harmonic oscillator, which will be employed throughout this paper. We first consider 
the one-dimensional shifted harmonic oscillator (SHO) defined by its position dependent 
potential:
\begin{equation} \label{a6}
{V^{(1)}}(x) =\frac{1}{2} \left(x^2-1\right)
\end{equation}
where we adopted the units $\hbar=\omega=m=1$. In this case the ground state energy is shifted 
to zero, therefore the energy spectrum is given by $E_n^{(1)} = n$ with $n \geqslant 0$ and the corresponding normalized wave functions may be written in terms of the Hermite polynomials 
$H_n(x)$ as:
\begin{align} \label{a5}
\psi _n^{(1)}(x) & =\frac{e^{-\frac{x^2}{2}} H_n(x)}{\sqrt[4]{\pi } \sqrt{2^n n!}}  \\
|n\rangle & = \int\!\! dx\, \psi _n^{(1)}(x)\, |x\rangle
\end{align}
Within the techniques of supersymmetric quantum mechanics (SUSY QM) mentioned earlier in 
Section \ref{iso section}  and using the expression (\ref{a1}), one obtains the function 
$I(x)$ in term of error function $\text{Erf}(x)$:
\begin{equation} \label{a7}
    I(x)=\frac{1}{2} \left[1+ \text{Erf}(x)\right].
\end{equation}
Then, using Eqs. (\ref{a2}),  (\ref{a6}) and (\ref{a7}), the one-parameter family of isospectral 
potentials ${{\widehat{V}}^{(1)}(\lambda\,;x)}$ are given by
\begin{align} \label{a8}
{{\widehat V}^{(1)}(\lambda\,;x)} =\frac{1}{2} \left(-1+x^2 + \frac{4 \lambda ^2 e^{-2 x^2}}{\pi [1+\sqrt{2} \lambda   I(x)
]^2} +\sqrt{\frac{2}{\pi }} \frac{4  \lambda  e^{-x^2} x}{1+\sqrt{2} \lambda   I (x)} \right)
\end{align}
In the top panel of Fig.\ref{isoSHO}, we show the isospectral potentials ${{\widehat V}^{(1)}(\lambda\,;x)}$  as a function of the position coordinate $x$ for three different values of $\lambda$. We can see that when $\lambda$ increases,  the form of ${{\widehat V}^{(1)}(\lambda\,;x)}$ starts developing a local minimum, which becomes deeper and narrower as well as $\lambda\rightarrow+\infty$, whereas for $\lambda\rightarrow0$ one gets back the SHO potential, which itself is a member of the isospectral family.  As a consequence, the non-linearity of such oscillators is characterized by the deformation parameter $\lambda$ and  for $\lambda  >  - \frac{1}{{\sqrt {2} }}$ the isospectral deformed potentials ${{\widehat V}^{(1)}(\lambda\,;x)}$ has no singularities. For small values of $\lambda$ and close to the origin of the real axis we may write 
$$
{{\widehat V}^{(1)}(\lambda\,;x)} = \frac12(x^2-1) + 2 \sqrt{\frac{2}{\pi}} \lambda (x-x^3) + O(x^4)\,,
$$
which intuitively explains the presence of a {\em displaced} minimum, due to the additional linear term, and of the {\em distortion}, due to the nonlinear cubic terms, in the isospectral deformed potentials.

The ground state of these isospectral oscillators is detached from the rest of the eigenstates and it corresponds to a bound state with zero energy $ \widehat E_0^{(1)} =0$. Notice that the ground states become more localized and develop larger peaks for larger values of the deformation parameter $\lambda$. Using Eq. (\ref{a3}), the ground state normalized wave function can be obtained as follows
\begin{equation} 
\phi_0(x;\lambda) = \frac{e^{-\frac12 x^2}}{\pi^{\frac14}} \frac{\sqrt{1+\sqrt{2}\lambda}}{1+\sqrt{2} \lambda I(x)}\,.
\label{psi00}
\end{equation}
The behavior of $\phi_0(x;\lambda) $ for some values of $\lambda$ is shown in the upper-right panel of Fig. \ref{isoSHO}. For small values of the deformation parameter we have
\begin{equation} 
\phi_0(x;\lambda) = \frac{e^{-\frac12 x^2}}{\pi^{\frac14}} \left[
1 - \frac{\lambda}{\sqrt{2}}\, \text{Erf}(x)
\right]\,.
\label{psi00s}
\end{equation}

Besides, the excited eigenstates are given by
\begin{align}
\phi_{n+1}(x;\lambda) =&\frac{2^{-\frac{n}{2}-1} e^{-\frac{3}{2}  x^2} }{\pi ^{3/4}  [1 + \sqrt{2} \lambda   I(x)] \sqrt{(n+1)!}}\qquad\qquad n=0,1,... \notag\\
&\times\left[\sqrt{\pi } e^{x^2} (2\lambda   I (x) +\sqrt{2}) H_{n+1}(x)+2 \lambda  H_n(x)\right]\,. \\
|\phi_n\rangle = & \int\!\!dx\, \phi_{n}(x;\lambda)\, |x\rangle
\label{psi01}
\end{align}
The first-excited state wave functions are shown in the lower-right panel of Fig. \ref{isoSHO}, for some values of $\lambda$.
%%%
%%%
\section{Non-Gaussianity and non-classicality of stationary states} \label{s:nongnonc}
This section is devoted to the study of the properties of the ground and thermal states of the isospectral SHO. Our analysis begins by evaluating their non-Gaussianity  and we investigate their non-classicality.
\subsection{Ground states}
We start by analysing the one parameter family of ground states 
\begin{equation}
{{\rho _\lambda } = \left| \phi_0 \right\rangle \left\langle\phi_0\right|}
\end{equation}
The ground state is a pure state, hence its Von Neumann entropy is equal to zero $S\left( \rho_{\lambda} \right)=0$, therefore the nonG measure (\ref{nong}) reduces to the following form:
\begin{equation}
    \delta[\rho_{\lambda}]=h\left( {\sqrt {\det \,\sigma[\rho_{\lambda}]} } \right),
\end{equation}
For this, the covariance matrix $\sigma[\rho_{\lambda}]$ is  in a diagonal form, where diagonal elements are the variances of canonical operators $\sigma_{11}={{\Delta X}  ^{2}}$, $\sigma_{22}=\Delta P ^{2}$. The expectation values and hence the variances can be calculated as in  \ref{Apx-A}.

After looking at the shape of the isospectral ground states in Fig.(\ref{isoSHO}) one would expect that the nonG vanishes for $\lambda \to 0$ and increases with $\lambda$. Indeed, this intuitive behavior is 
captured by a numerical evaluation of $\delta \left[ \rho_{\lambda} \right]$ and the 
corresponding results are reported in the Fig.(\ref{GNONG}) below. 
%%%
\begin{figure}[h!]
\centering
\includegraphics[width=0.8\columnwidth]{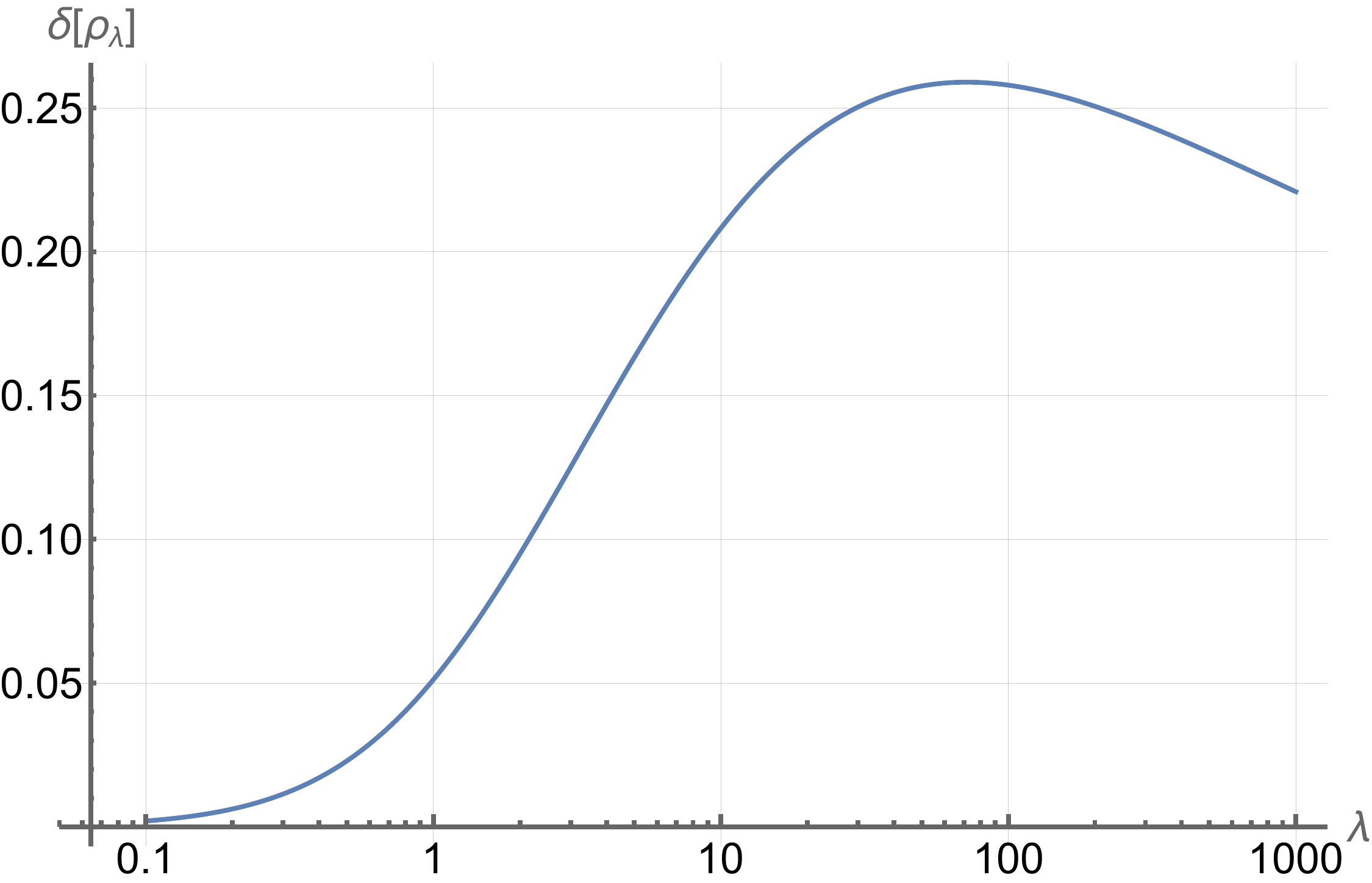}
\caption{QRE-based nonG $\delta \left[ \rho_{\lambda} \right]$ for the isospectral ground states $\rho_{\lambda}$ as a function of the deformation parameter $\lambda$.}
\label{GNONG}
\end{figure}
%%%

The nonG increases monotonically (roughly $\propto \lambda^2$) for lower values of the deformation parameter and grows to reach a maximum value $\delta \left[ \rho_{\lambda} \right] \simeq 0.26$  for $\lambda \simeq 71$ and then, for larger values of $\lambda$, it starts to slowly decrease. This does 
not seem to have an intuitive explanation, and in order to bettr capture the non-linear features 
of the deformed potentials we will also look, in the next Section, at the non-Gaussian 
properties of thermal states, 
which account for the whole spectrum of eigenstates. We conclude that the SUSY procedure has 
successfully generated an isospectral potentials with eigenstates having the desired 
non-Gaussian features.

\begin{figure}[h]
\centering
\includegraphics[width=0.8\columnwidth]{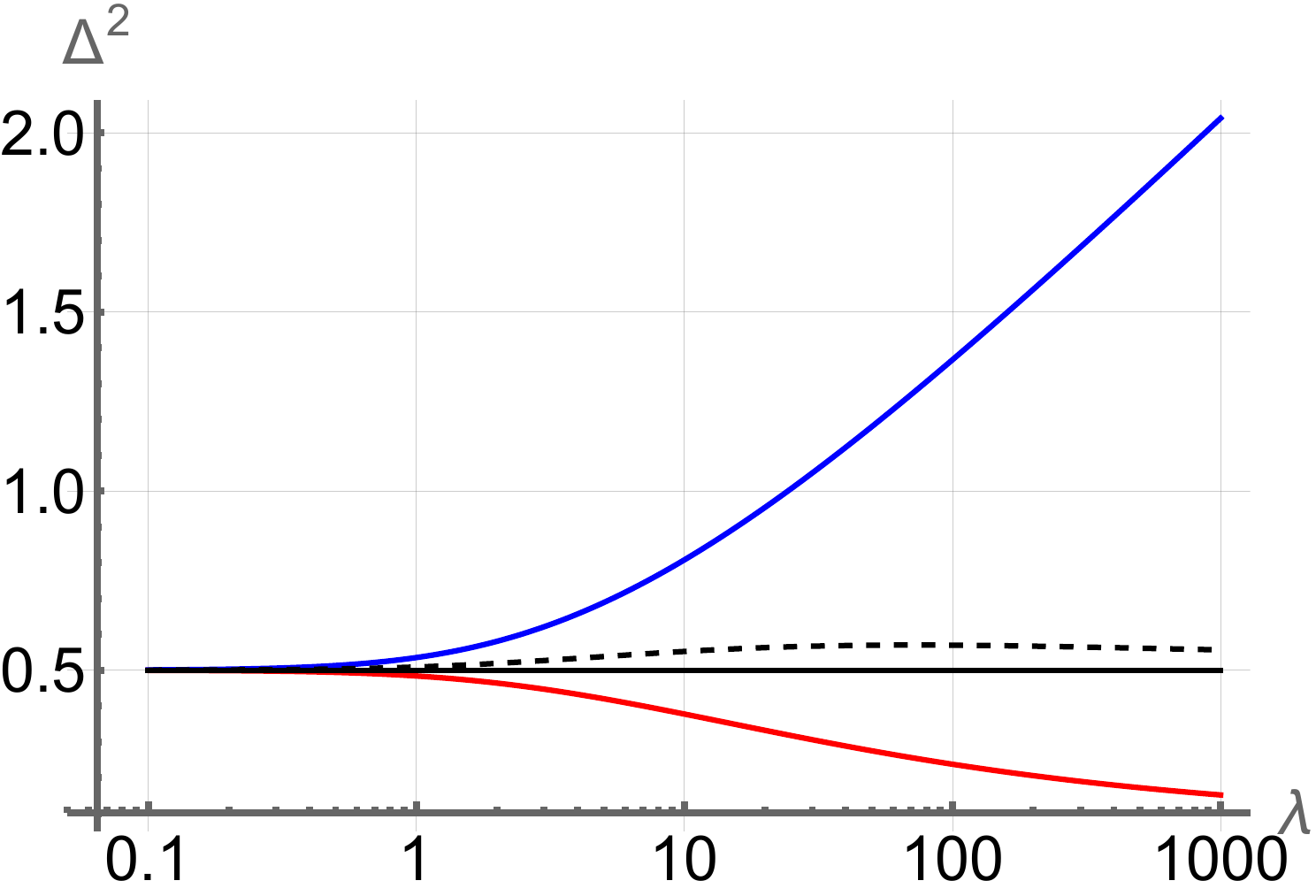}
\caption{The variances $\Delta X^2$ (red) and
$\Delta P^2$ (blue) for the isospectral ground states $\rho_{\lambda}$ as a function of the deformation parameter $\lambda$. The solid black line is the shot noise limit $\Delta X^2=\frac12$ and the dashed line denotes the uncertainty product $\Delta X \Delta P$, showing that the isospectral ground state is no longer a minimum uncertainty state. }
\label{SqzFig}
\end{figure}

\begin{figure}[h!]
\includegraphics[width=0.49\columnwidth]{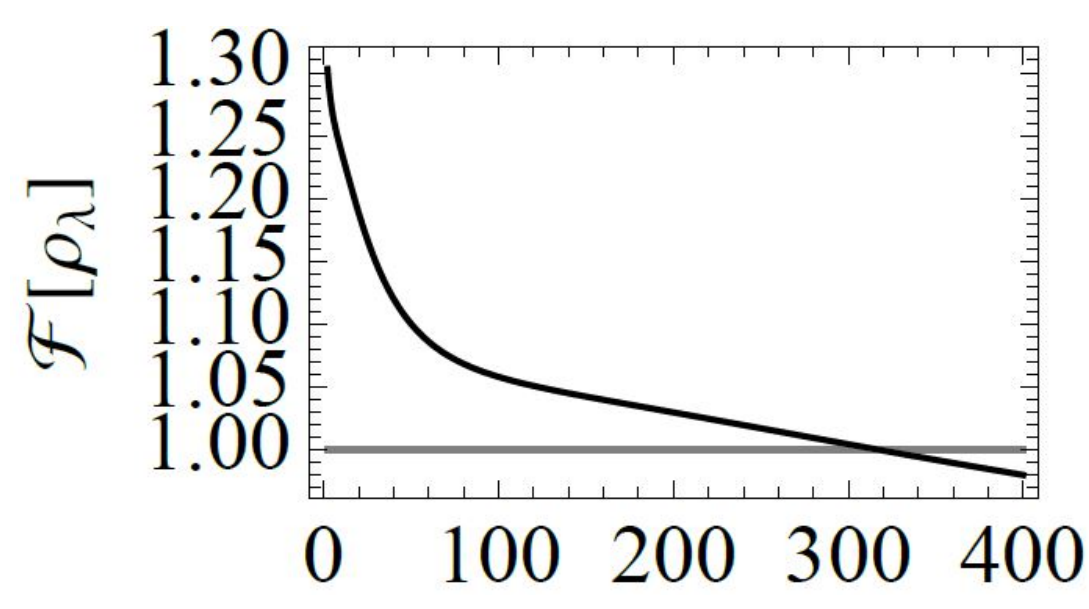}
\includegraphics[width=0.49\columnwidth]{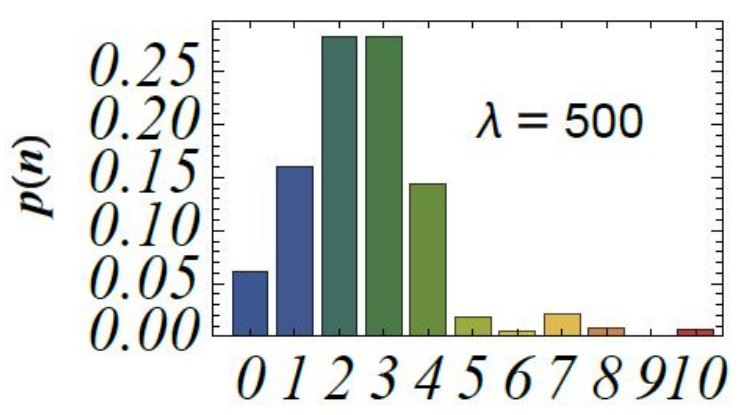}
\includegraphics[width=0.49\columnwidth]{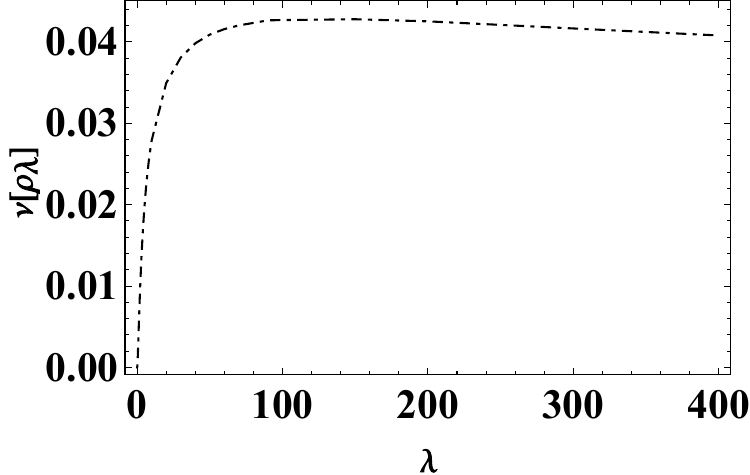}
\includegraphics[width=0.49\columnwidth]{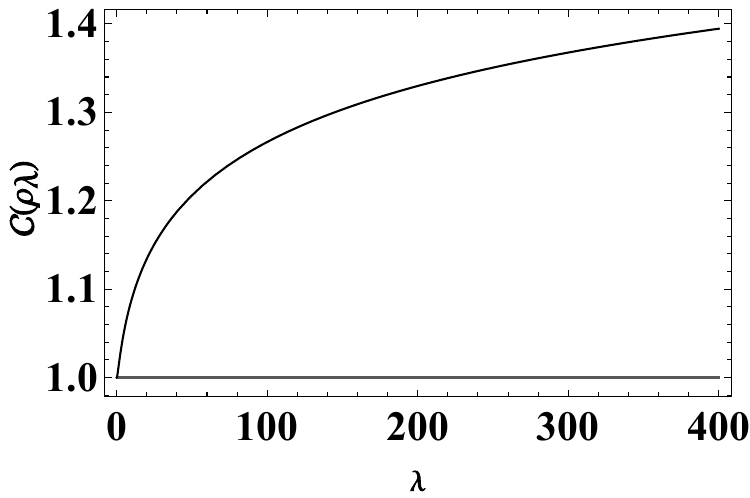}
\caption{
(Top left: Fano factor ${\mathcal{F}}\left[\rho_{\lambda}\right]$ for the states $\rho_{\lambda}$ as a function of the parameter $\lambda$. Top right: the inset displays the photon number distribution $p(n)$ for the ground state $\rho_{\lambda}$ , with $\lambda=500$.
(Bottom Left): Wigner negativity $\nu[\rho_{\lambda}]$ for the isospectral ground states $\rho_{\lambda}$ as a function of the deformation parameter $\lambda$. (Bottom right): Quadrature coherence scale ${\cal C}\left( \rho_{\lambda} \right)$ for the states $\rho_{\lambda}$ as a function of the parameter ${\lambda}$.}
\label{GNONC}
\end{figure}

The ground state is also a nonclassical state, as it may be shown by evaluating the variances 
of the quadrature operators $X$ and $P$. In particular, the ground state $\rho_{\lambda}$ 
exhibits squeezing, see Fig.\ref{SqzFig}), in the   position quadrature (red line) for 
any value of the deformation parameter. For $\lambda\rightarrow 0$, both variances go back 
to the shot noise limit (black line). Notice that the ground states of the isospectral potentials 
are no longer minimum uncertainty states (the dashed line denotes the uncertainty product).  
%%%%%
%%%%
At variance with quadrature squeezing, the isospectral ground states exhibit photon number squeezing 
only for large values of the deformation parameter. As it is apparent from 
the left upper panel of Fig. \ref{GNONC}, the Fano factor is larger than one
for $\lambda \leqslant 315$ and we have a subPoissonian distribution only 
for larger values of $\lambda$. An example of photon distribution is shown in the 
upper right panel of the same figure. 

Finally, in order to quantify the non-classicality of the ground states family, we 
have numerically evaluate both the Wigner negativity function (\ref{NC}) and the 
quadrature scale variance (\ref{QCS}). The results are reported in the lower 
panels of Fig.\ref{GNONC}, respectively. Notice that for pure states the QCS 
reduces to the so-called total noise
\begin{equation}
{{\cal C}^2}\left( \rho_{\lambda}  \right) = \Delta X^2 + \Delta P^2
\label{GQCS}
\end{equation}

Looking at Fig.\ref{GNONC}, we see that W-nonclassicality measure $\nu[\rho_{\lambda}]$, in the left lower panel, shows a behaviour analogue to that
of the non-Gaussianity measure $\delta[\rho_{\lambda}]$ in Fig.\ref{GNONG}, whereas for the QCS, in the right panel, we can observe that it increases continuously with $\lambda$ and for $\lambda  \to 0$ it reaches its minimal value ${\cal C}\left( \rho  \right)=1$ corresponding to coherent states with 
Poissonian profile.
\subsection{Thermal states}
\label{tst}
Up to now we have handled just with ground states, i.e. we have assumed the system at zero 
temperature. In order to analyze more realistic situations, we now consider Gibbs state, 
i.e. states at thermal equilibrium at temperature $T$, and analyze how the non Gaussian 
and non Classical properties changes as a function of temperature.  

The one parameter family of single-mode deformed thermal states ${\rho _\lambda ^{th}}$ is made 
of mixtures of the eigenstates with Gibbs weights. In formula 

\begin{equation}
{\rho _\lambda ^{th}} = \sum\limits_{k = 0}^\infty {{p_k}(T)\left| \phi_k\right\rangle \left\langle \phi_k \right|}
\label{PsiTh}
\end{equation}
where 
\begin{equation}
p_k (T) = \frac{1}{Z} e^{-k/T}
\end{equation}
the Boltmann constant has been set to $k_{B}=1$ and $Z$ denotes the partition function
\begin{equation}
Z = \sum_{n=0}^\infty e^{-n/T} = \frac{e^{1/T}}{e^{1/T}-1}
\end{equation}
In the above formulas, we have already used the isospectrality of all the potentials. 

\begin{figure}[h]
    \centering
    \includegraphics[width=0.9\columnwidth]{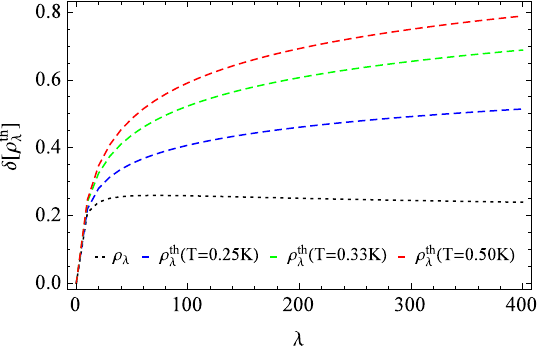}
    \caption{QRE based nonG $\delta \left[ {\rho _\lambda ^{th}} \right]$ as a function of the deformation parameter $\lambda$ for isospectral thermal states ${\rho _\lambda ^{th}}$ with $T=0.25 K$ (dashed  Blue), $T=0.33 K$ (dashed Green) and $T=0.50 K$ (dashed Red). The dotted Black line refers to the nonG of ground states family $\delta \left[ {\rho _\lambda}
     \right]$.}
    \label{TNONG}
\end{figure}

To appreciate the difference between the ground states and the thermal ones, in the following 
plots we also display the ground states value.  The Von Neumann entropy of the states 
${\rho _\lambda ^{th}}$ can be expressed as: 
\begin{equation}
    S\left( \rho _\lambda ^{th} \right) =  - \sum\limits_{k = 0}^\infty {{p_k(T)}\log {{p_k(T)}}} 
\end{equation}
hence, the QRE nonG measure (\ref{nong}) for the thermal states ${\rho _\lambda ^{th}}$ 
is given as follows 
\begin{equation}
\delta \left[ {\rho _\lambda ^{th}}  \right] = h\left( {\sqrt {\det \,\sigma \left[ {\rho _\lambda ^{th}}  \right] } } \right) + \sum\limits_{k = 0}^\infty {{p_k(T)}\log {{p_k(T)}}}
\end{equation}
In Fig.(\ref{TNONG}), we show the numerically evaluated QRE nonG measure for three thermal states ${\rho _\lambda ^{th}}$  with different values of temperature $T$ (see the legend). For comparison, we also show the nonG of the ground state.  For small values $\lambda \lesssim 10$, the QRE nonG measure increase monotonically with $\lambda$, exhibiting the same behaviour for all considered families of 
thermal states ${\rho _\lambda ^{th}}$. For larger values $\lambda > 10$, the nonG continues to 
increase, though less rapidly, and is larger for larger $T$. This means that temperature enhances 
the nonlinearity-induced nonGaussianity. The same, however, does not happen for non-classicality, 
as we discuss in the following.

\begin{figure}[h!]
  \includegraphics[width=0.49\columnwidth]{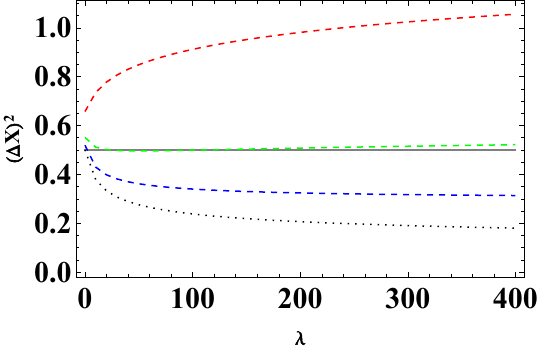}
  \includegraphics[width=0.49\columnwidth]{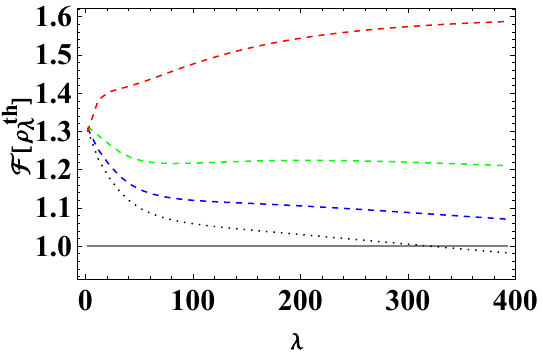}
  \includegraphics[width=0.49\columnwidth]{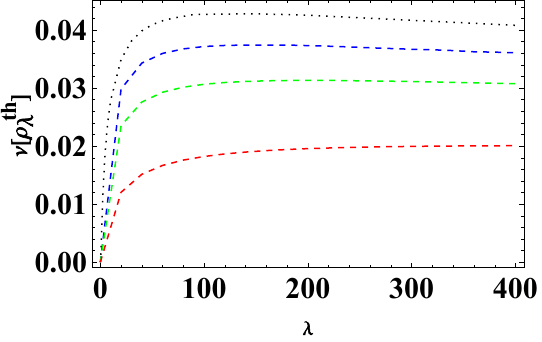}
  \includegraphics[width=0.49\columnwidth]{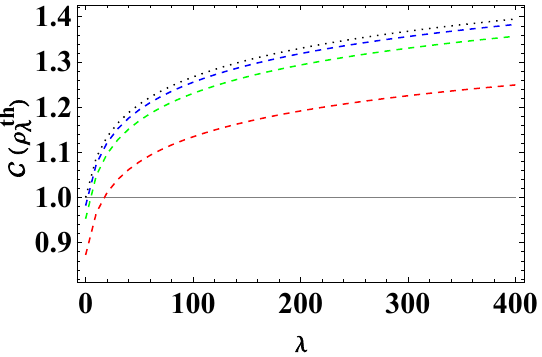}
  \caption{(Top left): Squeezing quadrature ${\left(  {\Delta X}\right)  ^{2}}$ for the isospectral thermal states ${\rho _\lambda ^{th}}$ as a function of the deformation parameter $\lambda$ for different value of the temperature. (Top right): Fano factor $\mathcal{F} \left[ {\rho _\lambda ^{th}} \right]$.(Bottom left): Wigner negativity $\nu[\rho_{\lambda}]$. (Bottom right): Quadrature coherence scale ${\cal C}\left( {\rho _\lambda ^{th}}  \right)$.
  In all plots, the red dotted line denotes the curve for $T =0.5K$, green is for $T =0.33K$ and blue for $T =0.25K$. The black dotted lines correspond to the measures of the ground states  ${\rho _\lambda}$, that is to the limit $T\rightarrow0$.}
\label{Squee}
\label{NONC01}
\end{figure}

From the evaluation of the variance  ${\left(  {\Delta X}\right)^{2}}$ for the three families of thermal states (see the upper left panel of Fig. \ref{Squee}), we can observe that thermal states exhibit squeezing 
for low temperatures $T<0.33°K$, while increasing temperature makes it to surpass the shot noise limit.
From the upper right panel of the same figure we see that for low temperatures $T<0.33°K$ the Fano factor decreases as the deformation parameter $\lambda$ increases. Considering the family of thermal states with $T=0.25°K$ (dashed Blue) we note a sub-Poissonian  distributions, $\mathcal{F} \left[ {\rho _\lambda ^{th}} \right]<1$ for $\lambda>900$. On the other hand, for higher temperatures $T\geq0.33°K$ the Fano factor increases with temperature. Finally, to further characterize and quantify the non-classicality of the families of thermal states, we have numerically evaluate both the Wigner negativity function (\ref{NC}) and the QCS (\ref{QCS}). Results are reported in the lower panels of Fig. \ref{NONC01}. Notice that QCS is calculated via the formula
\begin{equation}
{\cal C}^2 \left( \rho^{th}  \right) ={\cal P}^2 \left( \Delta X^2+ \Delta P ^2\right)
\end{equation}
where ${\cal P}$ is the purity. For these measures of non-classicality in thermal states, we see
a decrease for increasing temperature. Besides, at fixed value of $T$ they are affected by the 
deformation parameter $\lambda$. 
In particular, (see left panel) for low values of $\lambda$ the Wigner negativity increases monotonically as a function of $\lambda$ until reach a maximum, and then decreases slowly going to an asymptotic value. In the right panel we observe that QCS increases continuously with $\lambda$, whereas it approaches the limit ${\cal C}\left( \rho  \right)=1$ of \textit{"quadrature quasi-incoherent states"}  as $\lambda  \to 0$.

\section{Characterization of the SHO isospectral potentials}\label{s:qet}
Here we present some results about the characterization of isospectral oscillators, focussing on the ultimate bounds on the precision in estimating the deformation parameter $\lambda$. In particular, we address optimal measurement protocols 
and evaluate the corresponding quantum fisher information (QFI).

In the simplest configuration, the precision in the estimation of the deformation parameter 
$\lambda$ is quantified by the quantum Fisher information $H(\lambda)$ for the isospectral 
ground states $\rho_{\lambda}$, which itself may be evaluated using Eq. (\ref{QFIP}), 
arriving at  
\begin{equation}
H(\lambda)=\frac{2}{3} \frac1{( 1+ \sqrt 2 \lambda )^2}\,.
\end{equation}
Remarkably, the quantum Cramer-Rao bound is achieved by position measurements, i.e., by measuring 
the position operator $X$ of the oscillator, we have  $F(\lambda)=H(\lambda)$. The QFI is a decreasing function of  $\lambda$ and vanishes for great values of $\lambda$, thus indicating that any estimator $\widehat{\lambda}$ of the deformation parameter $\lambda$ becomes less and less precise for large $\lambda$. The QFI, however, decreases only quadratically with the parameter and therefore the signal-to-noise ratio scales as $\lambda^2/\hbox{Var}\lambda \rightarrow \frac13 - O(1/\lambda)$, i.e., it
remains finite even for $\lambda\rightarrow\infty$ %%%
For thermal states, the QFI for the parameter $\lambda$ at temperature $T$ is given by
\begin{align}
H(\lambda;T) =  2 &  \sum_{n\neq m}
\frac{[p_n(T)-p_m(T)]^2}{p_n(T)+p_m(T)} 
\left|\langle \phi_m | \partial_\lambda \phi_n \rangle \right|^2
\end{align}

\begin{figure}[h!]
\centering
\includegraphics[width=0.9\columnwidth]{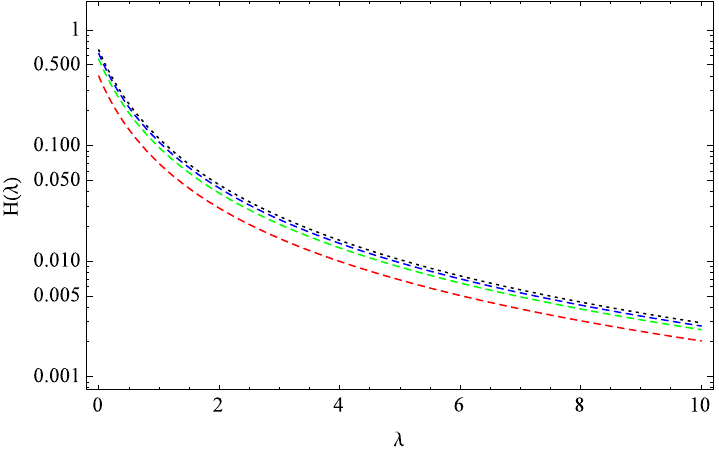}
    \caption{QFI for isospectral SHO thermal states ${\rho _\lambda ^{th}}$ as a function of the deformation parameter $\lambda$, The dashed curves are for increasing values of $T$ (shown in the legend) from top to bottom. the dotted line at the top is the ground state QFI.}
    \label{TQFI}
\end{figure}

In Fig.\ref{TQFI}, we show the behaviour of the quantum Fisher information $H(\lambda;T)$ at fixed temperatures as a function of $\lambda$. The QFI shows the same qualitative behaviour as that for 
ground states (Black dots), i.e. it decreases with the deformation parameter $\lambda$. At fixed $\lambda$, the QFI slightly decreases with temperature. Also in this case, the Fisher information of position measurement equals the QFI. In other words, position data provides us with all the maximum available 
information about the deformation parameter of the isospectral potentials.

\section{Conclusion}  \label{s:out}
Nonlinear oscillators  have attracted interest in different fields, as they play a relevant role for fundamental and practical purposes. In particular, they have recently received attention as a possible resource for quantum information processing. In this paper, we have focussed on SUSY potentials 
isospectral to the shifted harmonic oscillator (SHO) and shown that the ground state of the $\lambda$-isospectral potentials is non-Gaussian for all positive values of $\lambda$ with the non-Gaussianity persisting in thermal states and increasing with temperature. The states are also non-classical as confirmed by the emergence of squeezing of the ground state and also photon-number squeezing for large values of the deformation parameter. Additionally, we have shown that the ground state exhibits Wigner negativity, which increases with the deformation parameters until reaching a maximum and then slowly decreases. Non-classicality is also present in the equilibrium states of the systems up to some 
threshold temperature.

We have also addressed the characterization of the systems, i.e., the estimation of the the deformation parameter and evaluated the quantum Fisher information. We found for the pure state model the QFI decreases with increasing values of the parameter, and the Cramer-Rao bound is saturated by position measurement. The Quantum Fisher Information (QFI) for thermal states exhibits a qualitatively similar behavior with a slight decrease in value at higher temperatures. The QFI decreases for increasing $\lambda$ and vanishes in the limit 
$\lambda\rightarrow\infty$. The optimal measurement is still the position measurement which 
saturates the Cramer-Rao bound.

In conclusion, our study proves an interesting interplay between supersymmetric quantum mechanics and nonlinearity, unveiling the remarkable non-Gaussian and non-classical properties of the eigenstates of
$\lambda$-isospectral potentials. The persistence of these features in thermal states and their dependence on temperature and the deformation parameter confirm the robustness of these systems for exploring fundamental quantum phenomena.  Moreover, we have addressed the bounds to precision in the characterization of systems described by isospectral SUSY potentials, with position measurements emerging as an optimal strategy. 
Our results deepen our understanding of isospectral potentials and prove they may represent promising platforms for quantum information with continuous variables, particularly in leveraging their non-classical and non-Gaussian properties. We hope that our results will pave the way for further exploration, e.g., using coherent states of isospectral oscillators \cite{SANJAYKUMAR199673}, into the practical applications of nonlinearity and SUSY in quantum technologies.

\appendix

\section{Displacement vector and covariance matrix of a quantum state}
\label{Apx-A}
Introducing the real vector of canonical operators 
$R = {\left( X, P \right)^T}$, the displacement 
vector $E \left[ \rho  \right]$ and the covariance matrix $\sigma \left[ \rho  \right]$  of a quantum state are given by:
\begin{align}
E \left[ \rho  \right] & = \left( 
  { \left\langle X \right\rangle } , 
  {\left\langle {{P}} \right\rangle}  \right)^T
\end{align}
\begin{align}
\sigma \left[ \rho  \right] & = \left( 
\begin{array}{cc}
 \Delta X^2 & \frac{\langle XP\rangle + \langle PX\rangle}{2} - \langle X\rangle\langle P\rangle  \\ \frac{\langle XP\rangle + \langle PX\rangle}{2} - \langle X\rangle\langle P\rangle   & 
\Delta P^2
\end{array} \right)
\end{align}
For a generic pure state $|\phi\rangle$ 
we have
\begin{eqnarray}
\left\langle X \right\rangle    &= &  \langle \phi |X| \phi \rangle 
       = \int_\mathbb{R}\! dx\, x \,\left|\phi(x)\right|^2  \\
  \left\langle P \right\rangle   &= & \left\langle \phi |P| \phi \right\rangle   =  - i\hbar \int\limits_\mathbb{R}\! dx\, \phi(x) \,\partial_x\phi(x) \\
  \left\langle X^2\right\rangle   &= & \langle \phi|X^2| \phi \rangle   
  = \int_\mathbb{R}\!dx\, x^2 \,{\left| \phi(x) \right|^2} \\
  \left\langle P^2 \right\rangle   &= & \left\langle \phi \right|P^2\left| \phi\right\rangle  
  = -\hbar^2 \int_\mathbb{R}\! dx\, \phi(x) \,  \partial^2_x \phi(x)\\
  \left\langle XP \right\rangle  & = & \left\langle \phi \right|XP \left| \phi \right\rangle  
  = - i\hbar \int_\mathbb{R}\!dx \, \phi(x) \,x\, \partial_x \phi(x)   \\ 
  \left\langle PX\right\rangle  &= & \left\langle \phi \right|{P}X\left| \phi\right\rangle  
  = - i\hbar \int_\mathbb{R}\! dx\, \phi(x) \, \partial_ x [x\,\phi(x)] 
\end{eqnarray} 

\bibliographystyle{ws-ijqi}
\bibliography{isoho}
\end{document}